# An Advanced Features Extraction Module for Remote Sensing Image Super-Resolution

Naveed Sultan
Department of Electrical Engineering
Chulalongkorn University
Bangkok, Thailand
6570132321@student.chula.ac.th

Amir Hajian
Department of Electrical Engineering
Chulalongkorn University
Bangkok, Thailand
amirhajian85@gmail.com

Supavadee Aramvith
Multimedia Data Analytics and
Processing Research Unit
Department of Electrical Engineering
Chulalongkorn University
Bangkok, Thailand
supavadee.a@chula.ac.th

*Abstract*— In recent years, convolutional neural networks (CNNs) have achieved remarkable advancement in the field of remote sensing image super-resolution due to the complexity and variability of textures and structures in remote sensing images (RSIs), which often repeat in the same images but differ across others. Current deep learning-based super-resolution models focus less on high-frequency features, which leads to suboptimal performance in capturing contours, textures, and spatial information. State-of-the-art CNN-based methods now focus on the feature extraction of RSIs using attention mechanisms. However, these methods are still incapable of effectively identifying and utilizing key content attention signals in RSIs. To solve this problem, we proposed an advanced feature extraction module called Channel and Spatial Attention Feature Extraction (CSA-FE) for effectively extracting the features by using the channel and spatial attention incorporated with the standard vision transformer (ViT). The proposed method trained over the UCMerced dataset on scales 2, 3, and 4. The experimental results show that our proposed method helps the model focus on the specific channels and spatial locations containing high-frequency information so that the model can focus on relevant features and suppress irrelevant ones, which enhances the quality of super-resolved images. Our model achieved superior performance compared to various existing models.

## I. INTRODUCTION

Image super-resolution (ISR), a key component in the field of computer vision, focuses on enhancing low-resolution (LR) images to a high-resolution (HR) format. The significance of such high-resolution imagery lies in its detailed and fine-grained information, which is crucial across different sectors, including medical imaging, facial recognition, satellite images, text super-resolution, and video enhancement. Remote sensing images (RSIs) are used for disaster response, land cover analysis, object detection, and environmental surveys.

CNNs have become a primary method for improving the resolution of satellite imagery, offering a more efficient and cost-effective route to obtaining high-quality remote sensing data. Super-resolution (SR) methods are typically classified into three broad categories: interpolation-based [1], reconstruction-based [2], and learning-based [3]. Interpolation methods, known for their simplicity and efficiency, are widely used. These methods involve mapping target coordinates back to their original reference coordinates and then reconstructing the target pixel value based on the surrounding pixel value at the source. However, a notable drawback of interpolation-based methods is their general neglect of edge characteristics, often resulting in blurred edges in the enhanced images.

Most earlier satellite super-resolution (SR) techniques have relied on reconstruction-based methods. These methods reconstruct high-resolution (HR) images by utilizing subpixel information from low-resolution (LR) to multi-frame images [4]. Many reconstruction-based methods, such as sparse priors [5], incorporate prior knowledge to yield effective results. The limitation of these methods is reliance on feature design, which necessitates complex parameter tuning processes. This dependency makes it challenging. Deep learning methods have significantly progressed in remote sensing image super-resolution (RSISR). It is well known that restoring images with complex textures poses a significant challenge. Current SR techniques typically address this by sequentially stacking basic modules. CNN-based cutting-edge methods now focus on the feature extractions of RSIs using attention mechanisms. However, these methods are still not fully capable of identifying and utilizing key content attention signals in RSIs effectively:

1) Remote sensing images often contain valuable high-frequency details like edges and textures, which are crucial for accurate reconstruction. 2) poor feature extraction can lead to these details being missed or distorted, resulting in a blurry and inaccurate super-resolved image. It can distort the image's visual appearance and make it difficult to interpret the content accurately. 3) The inability to capture fine-scale spatial variations can affect tasks like identifying small objects and monitoring delicate environmental changes. To overcome these problems, we proposed an advanced feature extraction module called Channel and Spatial Attention-Feature Extraction (CSA-FE) that can significantly influence the model's performance. CSA-FE comprises a Channel Attention Module (CAM) and a Spatial Attention Module (SAM). Channel attention analyzes each channel's importance, highlighting informative ones with critical features like edges and textures.

In contrast, spatial attention analyzes specific locations within a channel, capturing fine-grained details and enhancing important spatial relationships, and a combination of both uses both mechanisms for richer feature representation, leading to sharper and clearer super-resolved images, in remote sensing. Then, this module is used for further processing by the transformer network ViT [6].

The standard ViT uses many encoders and decoders for processing. Which cause the problem of computational cost and simple feed-forward network (FFN) [7] that has a non-linear activation and two linear projections to extract the features. However, this approach ignores modeling spatial representation. Besides, the presence of redundant information on channels hinders feature expression competency. To overcome this problem, we employ a Spatial feed-forward network (SGFN) [8]. Our SGFN outperforms FFN by effectively capturing non-linear spatial information and reducing channel redundancy in fully connected layers. In addition, unlike earlier studies, the SG module employs depth-wise convolution to ensure computational efficiency.

## II. RELATED WORK

Up to now, many researchers have focused on remote sensing image resolution with state-of-the-art models. In this section, we present a comprehensive overview of the research on natural images SR and remote sensing images SR, focusing specifically on methods that employ attention-based, CNN-based, and Transformer-based approaches.

### A. Natural Images SR

In recent years, Image SR powered by deep learning has gained prominence because of the robust feature extraction competencies of the neural networks. This trend has accelerated the expansion and refinement of super-resolution algorithms. Dong et al. [9] pioneered deep neural networks for the SR problem. Their model, known as SRCNN, aimed to learn a mapping function directly from low-resolution to high-resolution images. SRCNN, demonstrating the potential of deeper networks in the SR domain.

Li et al. [10] developed the Multi-Scale Residual Network (MSRN), which utilizes adaptive feature extraction and leverages hierarchical information for image super-resolution (SR). The Deep Recurrent Fusion Network (DRFN) [11] also introduced an innovative transposed layer method for scale computation, enhancing the model's efficiency in dealing with various image scales. Moreover, the Information Distillation Network (IDN) [12] introduced a novel multiple-cascaded information distillation block designed to construct high-quality residuals for more accurate and efficient super-resolution processes. In image super-resolution (ISR), deep learning-based methods leverage significant challenges, particularly in striking a balance between computational efficiency and the quality of the resultant images. A notable example is the Squeeze-and-Excitation Next (SENext) [13]. This method effectively balances the performance with computational demands and reduces the overfitting risks. However, addressing these challenges thoroughly and effectively remains an ongoing area of exploration and focus.

### B. Remote Sensing Images SR

Spatial resolution is a critical metric for assessing the performance of remote-sensing images. HR remote-sensing images carry more information, which is highly beneficial for various remote-sensing tasks. This field of research has explored various methods for high-quality image enhancement, such as discrete wavelet transforms (DWTs) [14]. Drawing inspiration from networks created for super-resolution in natural images SR, Lei et al. [15] introduced the local–global combined Network (LGCNet). This network is specifically designed to learn multilevel representations of remote-sensing images, capturing both intricate local finer details and broader global environmental priors. A recent study uses Mixed higher-order attention [16], which consists of a concatenation block inside a concatenation group that adopts channel-wise concatenation followed by $1 \times 1$ convolution, and it has attained top-notch results.

Additionally, some researchers have been focusing on improving the representational capabilities of networks to more effectively extract multiscale features present in remote sensing (RS) images. For instance, Zhang et al. [17] introduced PRDNN, which is competent in progressively acquiring knowledge about satellite image characteristics across different levels and receptive areas. Wang et al. [18] developed AMFFN, which aimed at optimizing the effectiveness of information utilization and faster processing using residual deep neural networks. Huan et al. [19] proposed a novel Pyramidal Multiscale Residual Network (PMSRN), which enhances feature representation by fusing hierarchical features into a multiscale dilation residual block (MSDRB). Furthermore, Lei and Shi [20] have taken a unique approach by leveraging mixed-scale self-resemblance details in RS images, utilizing the global attention mechanisms. These innovations represent significant strides in the field, enhancing the ability to process and interpret complex RS imagery.

Models based on transformers have recently shown promise in various computer vision and NLP applications. Vaswani et al. [7] first proposed the attention-driven concept of transformers for use in machine translation. Instead of using a conventional recurrent neural network, which has difficulty encoding long-range connections between sequence pieces, transformers employ self-attention layers to capture these relationships. As seen in Fig. 1, vision transformers (ViTs) are introduced for the image identification job in [6]. These allow for the effective capturing of long-range dependencies within an input picture. ViTs [6] can decode an image into a series of patches and then process them. Over the last several years, the remote sensing community has investigated several ViT-based methods for diverse applications for natural language processing and images, especially in classification, identification, segmentation, and super-resolution. A recent article [21] presents a multi-stage enhancement framework based on transformers, combining features from several stages for different resolutions. The proposed approach involves using a multi-stage transformer structure in combination with the conventional super-resolution techniques. The reference [22] proposes a hybrid architecture combining a CNN and a transformer to include global and local feature details for super-resolution.

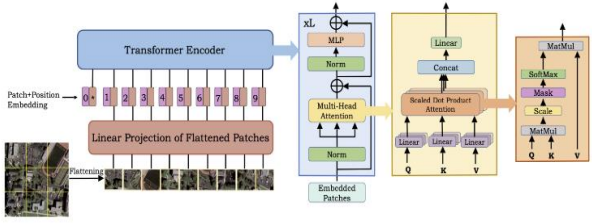

Fig. 1. The standard Vision Transformer network, and transformer encoder. Patches are first created from the input picture, then followed by flattening and projection onto a feature space. A transformer encoder is used to evaluate these features and provide the classification result.

The work [23] investigates the problem of multi-image super-resolution that combines many low-resolution remote sensing images. However, these transformers are still incapable of extracting and processing the high-level spatial information crucial in remote sensing.

### III. PROPOSED METHOD

In this section, we will explain the details of the proposed method, channel and spatial attention features extraction, and model diagram with a specific spatial gate feed-forward network. Fig. 2 presents an overview of our proposed architecture for remote sensing image super-resolution (RSISR). The Overall flow of the paper is Channel and Spatial attention features extraction, which is discussed in Section III-A, and the Transformer model, which is explained in Section II-B.

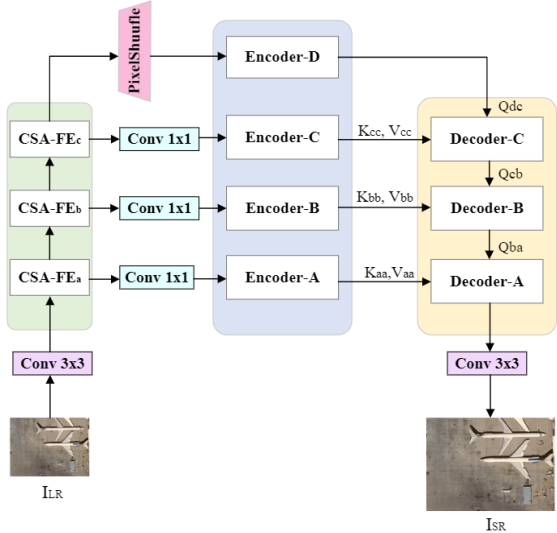

Fig. 1. Diagram of the proposed method

#### A. Channel and Spatial Attention Module

In the feature extraction module, we use channel and spatial attention, as shown in Fig. 3, Channel attention mechanisms seek to emphasize significant channels while diminishing less relevant ones. Channel attention takes the input features map F and utilizes max pooling $MP_c$ and average pooling $AP_c$ operations to reduce the dimensionality of data effectively and improve computational efficiency. We use reduction ratio to reduce the parameters using reduction ratio 16. The channel feature relationship between each feature map for each channel c is determined using a full convolution FC as shared dense layers in eq. 1.

$$F_c = FC\left(ReLU\left(FC\left(MP_c + AP_c\right)\right)\right) \quad (1)$$

Subsequently, an attention map is generated by summarizing the two outputs and applying the sigmoid function on $F_c$ for each channel c. The overall operations of channel attention are shown in eq. 2.

$$M_c = \sigma(F_c) \quad (2)$$

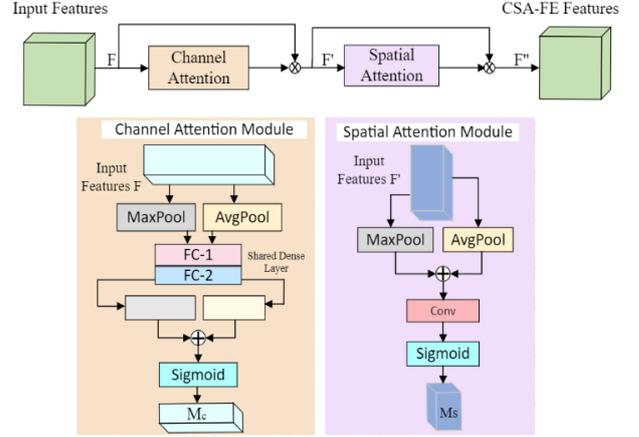

Fig. 2. Channel and Spatial Attention Module

Spatial attention is concerned with capturing relationships between regions within an image. This is critical in remote sensing, where fine-grained spatial features are often required. Spatial attention generates a single-channel feature map via convolution. The kernel size for convolution is 7×7×7. In contrast to channel attention, this method calculates the relationship between pixels, and then the sigmoid function is applied. Operations performed in Spatial attention are in eq. 3.

$$M_s = \sigma\left(Conv_{h,w}\left(MP_{h,w} + AP_{h,w}\right)\right) \quad (3)$$

$M_c$ is the output of the channel attention module after performing the channel attention operation. We use the summation function to add $M_c$ with the original input features, and we count its overall output as F′ later, F′ is used as an input for the spatial attention module and performs spatial operations. In contrast, $M_s$ is the output of SAM, and then it combines with the F′ features. Lastly, the output F′′ counted as CSA features for further processing. Mathematical operations are given in eq. 4.

$$F'_{c,h,w} = M_c \cdot F_{c,h,w} \quad \text{and} \quad F''_{c,h,w} = M_s \cdot F'_{c,h,w} \quad (4)$$

#### B. Vision Transformer

This section will discuss the vision transformer employed from [21] for the remote sensing image super-resolution. As shown in Fig. 2, this vision transformer uses three features extraction modules, as we discussed before, using the channel and spatial attention mechanisms. After feature extraction, we use an up sampler using a subpixel layer to convert they features from CSA-FE$_c$ low-dimension space to high-dimension space.

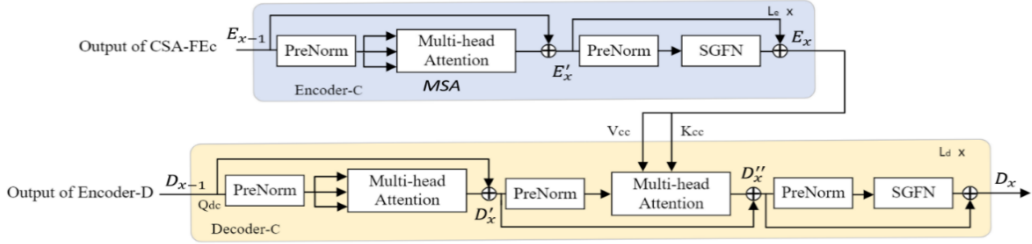

Fig. 3. Transformer Encoder and Decoder with Fusion Stage

This section will discuss the vision transformer employed from [21] for the remote sensing image super-resolution. As shown in Fig. 2, this vision transformer uses three features extraction modules, as we discussed before, using the channel and spatial attention mechanisms. After feature extraction, we use an up sampler using a subpixel layer to convert they features from CSA-FEc low-dimension space to high-dimension space. Further, many encoders and decoders are used for long-range dependencies and feature enhancement.. The feature dimensions are reduced by 1×1 convolution, and only one convolution is used to get a highly resolved image ISR. The model is trained using the L1 loss function. We need an SR image (ISR) and its associated HR reference (IHR) to calculate the loss; eq. 5 denotes the loss function.

$$L(\theta) = \frac{1}{N}\sum_{x=1}^{N}\|I_{HR}^{(x)} - I_{SR}^{(x)}\|_1 \quad (5)$$

where θ is the learnable parameter of the model, and N donates the iteration count.

*1) Encoder Block:* The transformer-based model accepts the series of 1-D tokens as an input, but to manage the 3-D features, the feature $f$ has $C \times H \times W$ as dimension divided into patches. Where C is the number of channels in the feature map, H is the height, and W is the width. These patches then transform into vectors $f_{px}$ having $P_h P_w C$ dimension, now $P_h$ and $P_w$, donates the height and width of the patch, and $x$ belongs from 1 to N. This approach adapts the transformer to work with three-dimensional data. The input of the transformer encoder is defined as eq. 6.

$$E_0 = [f_{p1}W, f_{p2}W, \ldots f_{pN}W] \quad (6)$$

W is a linear projection matrix with dimension $(P_h P_w C) \times D$. The encoder module contains multi-head attention (MSA) and spatial gate feed-forward network (SGFN) [8] We will discuss SGFN further below. MSA is the combination of several self-attentions. We use pre-layer normalization and residual structure before each module. The diagram in Fig. 4, represents a framework of encoder C. The remaining encoder and decoder have the same structure. The output of the encoder module is represented as

$$E'_x = MSA(PreNorm(E_{x-1})) + E_{x-1}, \quad x = 1\ldots,L_e$$
$$E_x = SFGN(PreNorm(E'_x)) + E'_x, \quad x = 1\ldots,L_e$$
$$[f_{E1}f_{E2}\ldots,f_{EN}] = EL_e \quad (7)$$

where $f_{Ex}$ is the output of the encoder module. we use decoders to incorporate the encoded representations of the features from each step of the proposed model.

*2) Decoder Block:* The decoder has the same structure as encoder multi-head attention and SGFN, but with one addition, a specific MSA block with cross attention from the encoder. It can manage the decoder input sequence and output of the prior encoder, an important part of the decoder function. The decoder output is computed as

$$D_0 = [f_{E1}f_{E2}\ldots f_{EN}]$$
$$D'_x = MSA(PreNorm(D_{x-1})) + D_{x-1}, \quad x = 1\ldots,L_d$$
$$D''_x = MSA(PreNorm(D'_x)PreNorm(D_0)) + D'_x, \quad x = 1\ldots,L_d$$
$$D_x = SFGN(PreNorm(D''_x)) + D''_x, \quad x = 1\ldots,L_d$$
$$[f_{D1}f_{D2}\ldots,f_{DN}] = DL_d \quad (8)$$

where $f_{Dx}$ is the decoder output and $L_d$ is the number of decoder layers. As aforementioned, a new approach we employ in our model is a Spatial gate feed-forward network (SGFN) [8] that uses the spatial gate to the FFN as shown in Fig. 5. It consists of depth-wise convolution and element-wise multiplication. It split the feature map into two main components: channel dimension for convolutional and multiplicative bypass. The overall input $X_{inp}$ SGFN is computed as

$$X' = \sigma(L_{1p}X_{inp}) \quad X' = X'_1 X'_2$$
$$SGFN(X_{inp}) = L_{2p}(X'_1 \odot (DW_c X'_2)) \quad (9)$$

where $L_{1p}$ and $L_{2p}$ represent linear projection, $\sigma$ is a GeLU function, and $DW_c$ is a learnable parameter for the depth-wise convolution. SGFN outperforms FFN by effectively capturing non-linear spatial information and reducing channel redundancy in fully connected layers. In contrast to earlier studies [24], [25], the SG module employs depth-wise convolution to ensure computational efficiency.

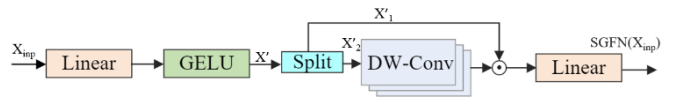

Fig. 4. Illustration of spatial feed-forward network (SGFN)

## IV. EXPERIMENTAL DETAILS

### A. lmplementation Details

This paper focuses on three scale factors, ×2, ×3, and ×4, for remote sensing images. We use PyTorch 2.0 and Nvidia GeForce GTX 180Ti GPU for training. Adam optimizer is used for optimization where β1 is 0.9, and β2 is 0.99. The initial learning rate is set to $1\times 10^{-4}$. A total 1500 epochs run to train the model, and batch size adjusted according to scale 10 for ×2, 8 for ×3, and 6 for ×4.

TABLE I. Mean Psnr and SSIM for the Ucmerced Test Dataset

| Scale | Matric | Bicubic [1] | SC [5] | SRCNN [9] | FSRCNN [26] | LGCNet [15] | DCM [27] | TransENet [21] | Ours |
|---|---|---|---|---|---|---|---|---|---|
| 2 | PSNR | 30.76 | 32.77 | 32.84 | 33.18 | 33.48 | 33.65 | 34.03 | **34.22** |
|   | SSIM | 0.8789 | 0.9166 | 0.9152 | 0.9196 | 0.9235 | 0.9274 | 0.9301 | **0.9320** |
| 3 | PSNR | 27.46 | 28.26 | 28.66 | 29.09 | 29.28 | 29.52 | 29.92 | **30.10** |
|   | SSIM | 0.7631 | 0.7971 | 0.8038 | 0.8167 | 0.8238 | 0.8394 | 0.8408 | **0.8439** |
| 4 | PSNR | 25.65 | 26.51 | 26.78 | 26.93 | 27.02 | 27.22 | 27.77 | **27.81** |
|   | SSIM | 0.6725 | 0.7152 | 0.7219 | 0.7267 | 0.7333 | 0.7528 | 0.7630 | **0.7645** |

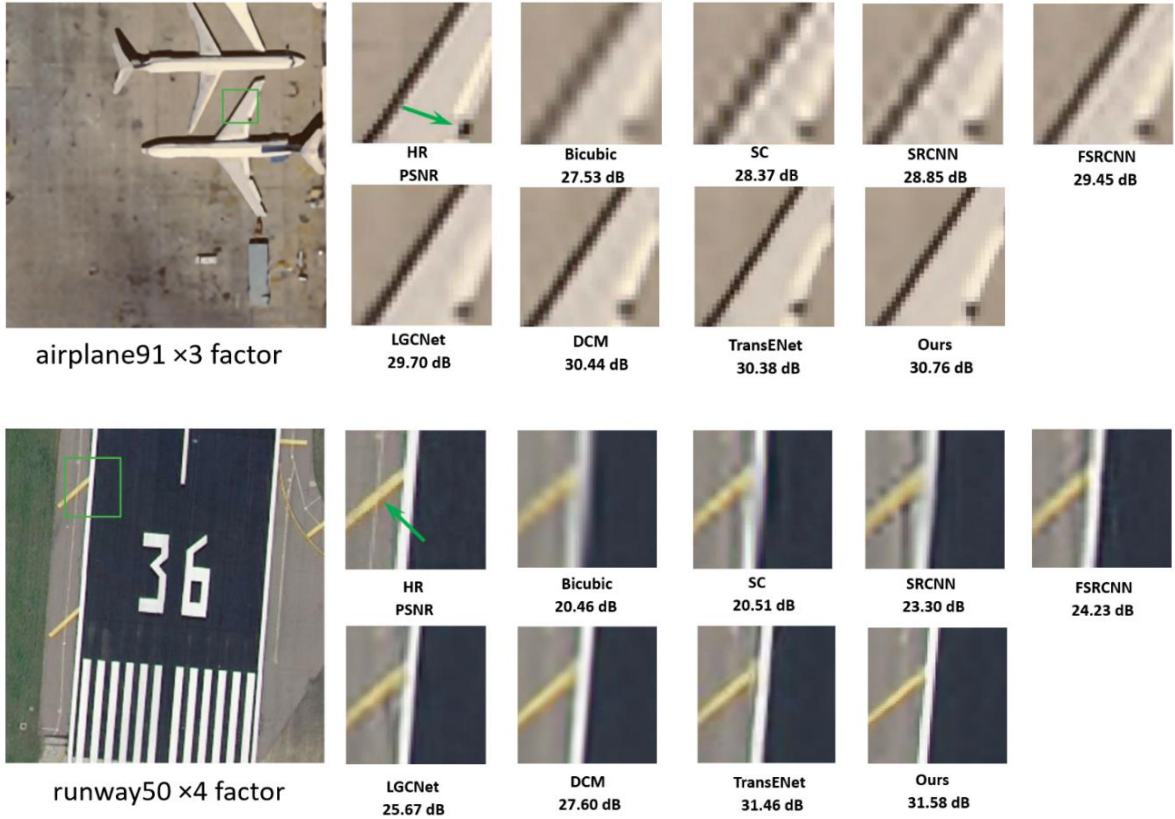

Fig. 5. Comparison with different methods, airplane91 with ×3, and runway50 with ×4

### B. Dataset

We use the public remote sensing dataset UCMerced [28]. It comprises 21 classes: agriculture, airplane, forest, harbor, rivers, etc. For each class 100 images, with $256 \times 256$ size. We split the dataset into two equal parts, 1050 images for training and 1050 images for testing, where 20% of the training is taken as validation.

### C. Quantitative Results

In our experiment, the original image was taken as an HR reference, and the LR image was acquired by bicubic interpolation. The quantitative results were measured in terms of peak signal-to-noise ratio (PSNR) and structural similarity index measure (SSIM) as shown in Table I. We evaluate our model by comparing with bicubic interpolation [1], sparse coding (SC) [5], SRCNN [9], FSRCNN [26], LGCNet [15], DCM [27] and TransENet [21], the best results are highlighted in red. The second-best is in blue Furthermore, we also compare our model for each class on × 3 shown in Table II. As it's clear from the table, our model performs well in most of the scenes with rich edges, texture, and spatial details.

TABLE II. Mean Psnr of Each Class for × 3 over Ucmerced Test Dataset

| class | Bicubic [1] | SC [5] | SRCNN [9] | FSRCNN [26] | LGCNet [15] | DCM [27] | TransENet [21] | Ours |
|---|---|---|---|---|---|---|---|---|
| 1 | 26.86 | 27.23 | 27.47 | 27.61 | 27.66 | 29.06 | 28.02 | 27.90 |
| 2 | 26.71 | 27.67 | 28.24 | 28.98 | 29.12 | 30.77 | 29.94 | 30.28 |
| 3 | 33.33 | 34.06 | 34.33 | 34.64 | 34.72 | 33.76 | 35.04 | 35.16 |
| 4 | 36.14 | 36.87 | 37.00 | 37.21 | 37.37 | 36.38 | 37.53 | 37.63 |
| 5 | 25.09 | 26.11 | 26.84 | 27.50 | 27.81 | 28.51 | 28.81 | 29.15 |
| 6 | 25.21 | 25.82 | 26.11 | 26.21 | 26.39 | 26.81 | 26.69 | 26.77 |
| 7 | 25.76 | 26.75 | 27.41 | 28.02 | 28.25 | 28.79 | 29.11 | 29.34 |
| 8 | 27.53 | 28.09 | 28.24 | 28.35 | 28.44 | 28.16 | 28.59 | 28.60 |
| 9 | 27.36 | 28.28 | 28.69 | 29.27 | 29.52 | 30.45 | 30.38 | 30.76 |
| 10 | 35.21 | 35.92 | 36.15 | 36.43 | 36.51 | 34.43 | 36.68 | 36.74 |
| 11 | 21.25 | 22.11 | 22.82 | 23.29 | 23.63 | 26.55 | 24.72 | 25.12 |
| 12 | 26.48 | 27.20 | 27.67 | 28.06 | 28.29 | 29.28 | 29.03 | 29.24 |
| 13 | 25.68 | 26.54 | 27.06 | 27.58 | 27.76 | 27.21 | 28.47 | 28.66 |
| 14 | 22.25 | 23.25 | 23.89 | 24.34 | 24.59 | 26.05 | 25.64 | 25.98 |
| 15 | 24.59 | 25.30 | 25.65 | 26.53 | 26.58 | 27.77 | 27.83 | 28.16 |
| 16 | 21.75 | 22.59 | 23.11 | 23.34 | 23.69 | 24.95 | 24.45 | 24.90 |
| 17 | 28.12 | 28.71 | 28.89 | 29.07 | 29.12 | 28.89 | 29.25 | 29.23 |
| 18 | 29.30 | 30.25 | 30.61 | 31.01 | 31.15 | 32.53 | 31.25 | 31.39 |
| 19 | 28.34 | 29.33 | 29.40 | 30.23 | 30.53 | 29.81 | 31.57 | 31.69 |
| 20 | 29.97 | 30.86 | 31.33 | 31.92 | 32.17 | 29.02 | 32.71 | 32.91 |
| 21 | 29.75 | 30.62 | 30.98 | 31.34 | 31.58 | 30.76 | 32.51 | 32.74 |
| AVG | 27.46 | 28.23 | 28.66 | 29.09 | 29.28 | 29.52 | 29.92 | 30.11 |

*D. Qualitative Results:*

We compare the results with different models shown above in Fig. 6 for $\times 3$ and $\times 4$. We can see by visual comparison that our model obtains better results.

## V. CONCLUSION

In this paper, our proposed channel and spatial attention feature extraction (CSA-FE) emphasizes significant channels in an image and captures relationships between regions where fine-grained spatial features are important. This helps the model to extract the features with rich edges, texture, and contours. Then, the encoder embeds the hierarchical features in the feature extraction, and the decoder fuses these encoded features to reconstruct the super-resolved image. The employed SGFN performs well to capture non-linear spatial information in both the encoder and decoder. Experimental results verify the effectiveness of our proposed model with 0.19, 0.18 and 0.04 dB on $\times 2$, $\times 3$, and $\times 4$, respectively compared to previous model. Our model has shown visually better quality of reconstructed image.

## ACKNOWLEDGEMENT

The research is supported in part by the Multimedia Data Analytics and Processing Research Unit, Department of Electrical Engineering, Chulalongkorn University, Bangkok 10330, Thailand, and the Graduate Scholarship Programme for ASEAN or Non–ASEAN Countries.